\documentclass[12pt,epsfig]{article} 
\input{epsf.tex}
\usepackage{amssymb,epsfig}
\newcommand{\One}{1\kern-4.5pt1}
\newcommand{\lapprox}{\raisebox{-0.5ex}{$\ 
\stackrel{\textstyle<}{\textstyle\sim}\ $}}
\newcommand{\gapprox}{\raisebox{-0.5ex}{$\ 
\stackrel{\textstyle>}{\textstyle\sim}\ $}}
\begin{document}

\addtolength{\baselineskip}{0.20\baselineskip}

\rightline{SWAT/04/398}

\hfill June 2004

\vspace{48pt}

\centerline{\Huge Zero Lattice Sound}

\vspace{18pt}

\centerline{\bf Simon Hands$^a$ and
Costas G. Strouthos$^b$}

\vspace{15pt}

\centerline{$^a$ {\sl Department of Physics, University of Wales Swansea,}}
\centerline{\sl Singleton Park, Swansea SA2 8PP, U.K.}
\smallskip
\centerline{$^b$ {\sl 
Division of Science and Engineering, Frederick Institute of
Technology,}}
\centerline{\sl Nicosia 1303, Cyprus.}
\vspace{24pt}


\centerline{{\bf Abstract}}

\noindent
{\narrower
We study the $N_f$-flavor 
Gross-Neveu model in 2+1 dimensions with a baryon chemical
potential $\mu$, using both analytical and numerical methods.  
In particular, we study the self-consistent Boltzmann equation in the
Fermi liquid framework using the
quasiparticle interaction calculated to $O(1/N_f)$, and find solutions for zero
sound propagation for almost all $\mu>\mu_c$, the critical chemical potential
for chiral symmetry restoration. Next we present results of a numerical
lattice simulation, examining temporal correlation functions of mesons defined
using a point-split interpolating operator, and finding evidence for 
phonon-like behaviour characterised by a linear dispersion relation in the long
wavelength limit. We argue that our results provide the first evidence for a
collective excitation in a lattice simulation.
}


\bigskip
\noindent
PACS: 11.10.Kk, 11.15.Ha, 11.15.Pg, 21.65.+f, 67.80Cx

\noindent
Keywords: Monte Carlo simulation,
 Fermi liquid, zero sound, phonon

\vfill
\newpage
\section{Introduction}

Such is the perceived difficulty of performing non-perturbative lattice
simulations of systems with non-zero baryon chemical potential 
$\mu$ that it is often
overlooked that, in those cases where simulations are possible, the information
extracted considerably exceeds that coming from better understood simulations
at non-zero temperature $T$. The reason is that for $\mu/T\gg1$ there is no
stringent limitation 
on the number of temporal lattice spacings $L_t$,
so that correlation functions in Euclidean time can be studied with good
resolution. These functions encode all information about the spectrum 
of the system under study, including energies and widths of all 
quasiparticle and collective excitations. For instance, in Two Color QCD
spectral studies have revealed a scalar isoscalar
Goldstone boson signalling the superfluid
nature of the high density state \cite{KSHM}, 
and evidence for an in-medium decrease in mass of the 
$\rho$-meson \cite{vector}. In the NJL model in 2+1$d$ the quasiparticle
dispersion relation $E(k)$ is smooth for $k\simeq k_F$ implying a 
well-defined Fermi surface, but anomalously slow temporal
decay in the
scalar isoscalar
diquark channel signals a critical state characteristic of a thin film
superfluid \cite{HM,HLM}. In NJL$_{3+1}$ by contrast, the scalar 
isoscalar diquark channel
has a Goldstone pole, and the quasiparticle dispersion is gapped, providing
evidence for superfluidity via orthodox BCS diquark pair condensation
\cite{HW}.

In refererence \cite{HKTS} we have studied a still simpler model, the 2+1$d$
Gross-Neveu (GN) model describing $N_f$ flavors of self-interacting fermions, 
with continuum Lagrangian 
\begin{equation}
{\cal L}=\bar\psi(\partial{\!\!\!/\,}+\mu\gamma_0)\psi
-{g^2\over{2N_f}}(\bar\psi\psi)^2.
\end{equation}
As $\mu$ is increased, the model exhibits a sharp first-order transition 
at a critical $\mu_c$
from a phase where a fermion mass $\Sigma_0$
is dynamically generated but baryon density
$n_B=\langle\bar\psi\gamma_0\psi\rangle$ vanishes to one where $\Sigma_0=0$
and $n_B\propto\mu^2$ \cite{Simon}. 
By examining various temporal correlation functions we were able to 
demonstrate many features of this high density phase.
For example, 
the quasiparticle dispersion relation is again consistent
with a sharp Fermi surface, a feature reinforced by the oscillatory
behaviour of meson wavefunctions 
as the spatial separation between quark and antiquark fields is increased,
reminiscent of a phenomenon in many body physics known as Friedel oscillation.
Secondly, we studied the correlator of the scalar meson $\sigma$ and found that,
in contrast to its behaviour in vacuum, in the high density phase it decays 
with a pole mass $M_\sigma\simeq2\sqrt{\mu(\mu-\mu_c)}$, the value predicted 
to leading non-trivial order in $1/N_f$. Finally, we studied meson correlators
at non-zero momentum
with several different quantum numbers, and found that in general they decay
algebraically rather than exponentially with Euclidean time, since 
it is usually possible to excite a particle - hole pair at 
the Fermi surface with zero energy cost.

There is a phenomenological model of interacting degenerate systems, originally
due to Landau \cite{Landau}, known as the Fermi liquid. From a particle 
physics perspective it is conceptually simple; the dominant low energy
excitations, the so-called quasiparticles, carry the same quantum numbers (spin,
isospin, baryon number, \dots) as
free particles and holes, and can be viewed simply as dressed versions of the
elementary quanta. Beyond the explicit breaking of Lorentz symmetry introduced
by $\mu\not=0\Rightarrow n_B\not=0$, 
there is no broken symmetry or order parameter
which distinguishes the Fermi liquid from non-interacting degenerate fermions.
However if certain conditions are fulfilled, such as an
effective interaction between quasiparticles which is both short-ranged
and repulsive,
then there is a massless bosonic
excitation in the spectrum as $T\to0$. Carrying zero baryon
charge, it is a {\em phonon}, ie. a quantum of a collective
excitation called {\em zero sound} \cite{Landau4}. 
Sound propagation occurs in any elastic medium; zero sound happens when the
elasticity originates not from collisions between individual particles,
but from the force on a single particle due to its coherent
interaction with all others present in the medium.
It can be pictured as a propagating distortion 
in the local shape of the 
Fermi surface, whose speed exceeds that of conventional
``first'' sound.

In \cite{HKTS} we found one meson channel whose temporal decay resembled
that of an isolated pole rather than that of a particle - hole continuum, and
which moreover yielded a phonon-like dispersion relation $\omega\propto k$. 
In the
current paper we seek to build on our tentative identification of this state
with zero sound, firstly by finding an analytic solution appropriate for
GN$_{2+1}$, and secondly by performing a more refined numerical
analysis in the channel of interest. In Sec.~\ref{sec:background} we review
the derivation of the self-consistent equation for zero sound propagation in the
framework of orthodox Fermi liquid theory, and in Sec.~\ref{sec:theory} we 
solve this equation for the particular case of GN$_{2+1}$, using the $O(1/N_f)$
expression for quasiparticle interactions derived in \cite{HKTS}. We find
solutions for sound propagation for almost all $\mu>\mu_c$, with speed
$\beta_0>\beta_F$, the Fermi velocity, which is the characteristic scale for a 
degenerate system. In Sec.~\ref{sec:numbers} we present results of a numerical 
simulation of GN$_{2+1}$; we apply orthodox meson spectroscopy techniques and
find evidence for a state with dispersion $\omega(k)\propto k$ in the small-$k$
limit. The propagation speed is of the same order, but slightly less than, the
Fermi velocity. Our results are summarised with a discussion about possible
sources for the discrepancy in
Sec.~\ref{sec:discuss}.

\section{Theoretical Background}
\label{sec:background}

We begin with a brief introduction of the theory behind sound propagation in
degenerate systems, leaning heavily on the treatment in \cite{Landau2}.
Consider a system of (quasi)particles with phase space distribution $n$
given by
\begin{equation}
n(\vec k,\vec x,t)=n_0(\vec k)+\delta n(\vec k,\vec x,t)
\end{equation}
and energy 
\begin{equation}
\varepsilon(\vec k,\vec r,t)
=\varepsilon_0(\vec k)+\delta\varepsilon(\vec k,\vec r,t),
\end{equation}
where for states near the Fermi surface the equilibrium conditions are given by
\begin{eqnarray}
n_0(\vec k)&=&(\exp(\varepsilon_0(\vec k)-\mu)/T+1)^{-1}\nonumber\\
\varepsilon_0(\vec k)&\simeq&\mu+\beta_F(\vert\vec k\vert-k_F).
\end{eqnarray}
The parameters are Fermi momentum $k_F$, and Fermi velocity
$\beta_F\equiv\vert\vec\nabla_k\varepsilon_0(\vert\vec k\vert=k_F)\vert$, 
which in general differ from their free field theory values. Small departures
from equilibrium are described by
\begin{equation}
\delta\varepsilon(\vec k,\vec x,t)=\mbox{tr}\int{{d^dk^\prime}\over{(2\pi)^d}}
{\cal F}(\vec k,\vec k^\prime)\delta n(\vec k^\prime,\vec x,t),
\end{equation}
where $d$ is the number of spatial dimensions and
the trace is over internal degrees of freedom such as spin and flavor.
The Fermi liquid interaction ${\cal F}$ encodes the response of a single
quasiparticle state to a change in the many-body distribution, and is thus
non-zero only for interacting systems \cite{Landau}. To leading order in any
small expansion parameter, ${\cal F}$ is equivalent to the matrix element for
forward scattering \cite{Landau3}. A review of the Fermi liquid picture applied
to relativistic systems is given in \cite{BC}.

Departures from equilibrium are 
described by a transport (Boltzmann) equation
\begin{equation}
{{dn}\over{dt}}={{\partial\delta n}\over{\partial t}}+\vec\nabla\delta n.
\vec\nabla_k\varepsilon_0-\vec\nabla_k
n_0.\vec\nabla\delta\varepsilon=I(n),
\end{equation}
where we have used Hamilton's equations
\begin{equation}
{{\partial\vec x}\over{\partial t}}=\vec\nabla_k\varepsilon\;\;\;;\;\;\;
{{\partial\vec k}\over{\partial t}}=-\vec\nabla\varepsilon,
\end{equation}
and dropped terms of $O(\delta^2)$. The term $I(n)$ on the right hand side is
the collision integral describing the net number of scattering events
into the phase
space element $d^dxd^dk$ per unit time. For oscillation frequency $\omega$
with mean free time $\tau$ between quasiparticle collisions, 
the dimensionless parameter
$\omega\tau$ governs which kind of excitation dominates. For $\omega\tau\ll1$
collisions reestablish local thermodynamic equilibrium in each volume element,
and ordinary {\em first sound} waves propagate with velocity
$\beta_1=\sqrt{\partial p/\partial\varepsilon}\simeq k_F/\sqrt{d}\mu$. 
For $\omega\tau\gg1$,
collisions are unimportant, and local thermodynamic equilibrium no longer
holds. Since for a Fermi liquid $\tau\propto T^{-2}$, this situation holds
near temperature zero, and the consequent excitations are known as 
{\em zero sound}.

With $T=I=0$, we have
\begin{equation}
\vec\nabla_k\varepsilon_0=\beta_F\hat m\;\;\;;\;\;\;
\vec\nabla_k n_0=-\hat
m\beta_F\delta(\varepsilon-\mu)=-\hat m\delta(\vert\vec k\vert-k_F),
\end{equation}
where $\hat m.\vec k=\vert\vec k\vert$, $\hat m.\hat m=1$. 
Specialising to oscillatory excitations
at the Fermi surface of the form 
$\delta n=\delta(\varepsilon-\mu)\Phi(\hat m)
e^{i(\vec k.\vec x-\omega t)}$, we find the self-consistent zero sound
equation \cite{Landau4}
\begin{equation}
(\omega-\beta_F\hat m.\vec k)\Phi(\hat m)=\hat m.\vec k S_dk_F^{d-1}
\mbox{tr}\oint d\Omega
{\cal F}(\hat m,\hat m^\prime)\Phi(\hat m^\prime).
\label{eq:zs}
\end{equation}
where the integral over the unit sphere is
normalised to $\oint d\Omega=1$, and the geometrical factor
$S_d\equiv2/(4\pi)^{d\over2}\Gamma({d\over2})$.

\section{Analytic Solution}
\label{sec:theory}

For the GN$_{2+1}$ model considered here, the Fermi surface is 
a circle, and the Fermi liquid interaction a simple function of the angle
$\theta$ between quasiparticle momenta. 
Quasiparticle interactions in the GN model are mediated by a scalar
boson $\sigma$:
to leading order in $1/N_f$ in the
chiral limit the interaction arises from the exchange contribution to forward
scattering, is repulsive, and has been determined to be \cite{HKTS}
\begin{equation}
{\cal F}(\theta)={\pi\over{4N_f(\mu-\mu_c)}}(1-\cos\theta),
\label{eq:gnf}
\end{equation}
where $\mu_c$ is
the critical chemical potential at which the transition from chirally-broken
vacuum to chirally-symmetric quark matter takes place at $T=0$. To leading 
order in $1/N_f$ $\mu_c=\Sigma_0$, the dynamically-generated quark mass at
$T=\mu=0$. For reasons to be made clear below, we will generalise
(\ref{eq:gnf}) to read ${\cal F}\propto R-\cos\theta$ with
$R>1$ \footnote{ 
In principle this could arise through a small bare quark mass
$m$, whereupon the exchange term yields $R=1+m^2/\mu^2$ \cite{BC}. 
Note, though,
that for $m\not=0$ there is also an {\em attractive} contribution 
from the direct term, which moreover is
no longer $O(1/N_f)$.}. Using (\ref{eq:gnf}) in (\ref{eq:zs}) we derive
\begin{equation}
(\zeta-\cos\theta)\Phi(\theta)=G\cos\theta\oint{{d\theta^\prime}\over{2\pi}}
(R-\cos(\theta-\theta^\prime))\Phi(\theta^\prime),
\label{eq:gnzs}
\end{equation}
The variable $\zeta=s+i\gamma$ 
is the ratio of phase velocity $\beta_0\equiv\omega/\vert\vec
k\vert$ to Fermi velocity $\beta_F$.  The constant $G$ parametrising 
the strength
of the Fermi liquid interaction will be specified below.

To find a solution we attempt an ansatz of the form
\begin{equation}
\Phi(\theta)=\sum_n{{b_n\cos^n\theta+c_n\sin\theta\cos^n\theta}.
\over{\zeta-\cos\theta}}
\label{eq:phi}
\end{equation}
After substituting this into (\ref{eq:gnzs}) the integrals over $\theta^\prime$
can be performed using
\begin{equation}
\oint{{d\theta}\over{2\pi}}{{\cos n\theta}\over{\zeta-\cos\theta}}
=-{{2a^n}\over{a-a^{-1}}}\;\;\;;\;\;\;
\oint{{d\theta}\over{2\pi}}{{\sin\theta\cos n\theta}
\over{\zeta-\cos\theta}}=0
\label{eq:ints}
\end{equation}
where $a$, $a^{-1}$ are the roots of $z^2-2\zeta z+1=0$ and 
we specify $\vert a\vert\leq1$.
Equating coefficients of $\cos^n\theta$ it is straightforward to show 
that the only non-trivial coefficients in (\ref{eq:phi}) are
$b_1$ and $b_2$; we arrive at the coupled system
\begin{eqnarray}
b_1&=&-{{RG}\over{a-a^{-1}}}\left[2b_1a+b_2(1+a^2)\right],\nonumber\\
b_2&=&{G\over{a-a^{-1}}}\left[b_1(1+a^2)+{b_2\over2}(3a+a^3)\right]
\label{eq:simul}
\end{eqnarray}
and immediately deduce
\begin{equation}
{b_2\over b_1}={{a^{-2}-1-2GR}\over{GR(a+a^{-1})}}.
\label{eq:phib}
\end{equation}
Substituting this back into (\ref{eq:simul})
determines a cubic equation for $a^2$:
\begin{equation}
(a^2-1)(Ga^4+a^2(2RG^2+G(3-4R)-2)+2)=0.
\end{equation}
Discarding the trivial solution $a^2=1$ we find a regular real
solution with $a<1$ exists for $R>2/(2-G)$ and is given by
\begin{equation}
a^2={1\over G}\left(1+(2R-{\textstyle{3\over2}})G-G^2R-\sqrt{
(1+(2R-{\textstyle{3\over2}})G-G^2R)^2-2G}\right).
\end{equation}
The corresponding sound speed is given by
\begin{equation}
s={1\over2}(a+a^{-1})>1.
\end{equation}

Note that since $s>1$ the denominators of (\ref{eq:phi},\ref{eq:ints})
are never zero; there is no necessity to specify a pole prescription to
evaluate the integral and 
$\gamma=0$. Physically this means the wave is undamped.
There are no solutions of the form (\ref{eq:phi})
with complex $a$ which describe dissipative solutions 
with $s<1$ and $\gamma\not=0$.

We now return to the issue of why it is appropriate to consider $R>1$
for the GN model in the chiral limit. The expression (\ref{eq:gnf})
is derived from the relation ${\cal F}(\theta)=-{\cal A}(\theta)$, where
${\cal A}$ is the forward scattering amplitude for physical
quasiparticles on the Fermi surface, and is valid only in
the large-$N_f$ limit. 
Beyond this leading order it is necessary to consider a more
general amplitude ${\cal M}(\theta,q)$ with 
momentum transfer $q=(q_0,\vec q)$. 
For quasiparticle forward scattering at the Fermi surface
${\cal A}=\lim_{\vert\vec q\vert\to0}\lim_{q_0\to0}
{\cal M}(q)$,
whereas for the Fermi liquid interaction
${\cal F}=-\lim_{q_0\to0}\lim_{\vert\vec q\vert\to0}{\cal M}(q)$
\cite{Landau}
(see also \cite{AGDNO}).
Taking the non-commutativity of the limits into account
yields the relation
\begin{equation}
-{\cal A}(\theta)={\cal F}(\theta)+{{{\mathfrak g}k_F}\over{2\pi\beta_F}}
\oint{{d\theta^\prime}\over{2\pi}}{\cal A}(\theta^\prime){\cal
F}(\theta-\theta^\prime),
\end{equation}
where ${\mathfrak g}$ is the degeneracy of single particle states resulting
from the trace in (\ref{eq:zs}).
Expanding ${\cal A}(\theta)=-\sum_n P_n\cos n\theta$, 
${\cal F}(\theta)=\sum_n Q_n\cos n\theta$, we readily derive the general
relation
\begin{equation}
Q_n={P_n\over{1-{{{\mathfrak g}k_F}\over{4\pi\beta_F}}P_n}}.
\end{equation}
For the GN model the only non-vanishing coefficients are $n=0,1$; we 
recover (\ref{eq:gnzs}) with 
\begin{equation}
G={{2\tilde G}\over{2+\tilde G}};\;\;
\tilde G={{{\mathfrak g}k_F}\over{8N_f\beta_F(\mu-\mu_c)}}=
{{{\mathfrak g}\mu}\over{8N_f(\mu-\mu_c)}}+O(N_f^{-2});\;\;
R={{2+\tilde G}\over{2-\tilde G}}>{2\over{2-G}}.
\label{eq:G}
\end{equation}
A zero sound solution with $s>1$ therefore exists for all
$\mu\gapprox16N_f\mu_c/(16N_f-{\mathfrak g})$.

\begin{figure}[htb]
\bigskip\bigskip
\begin{center}
\epsfig{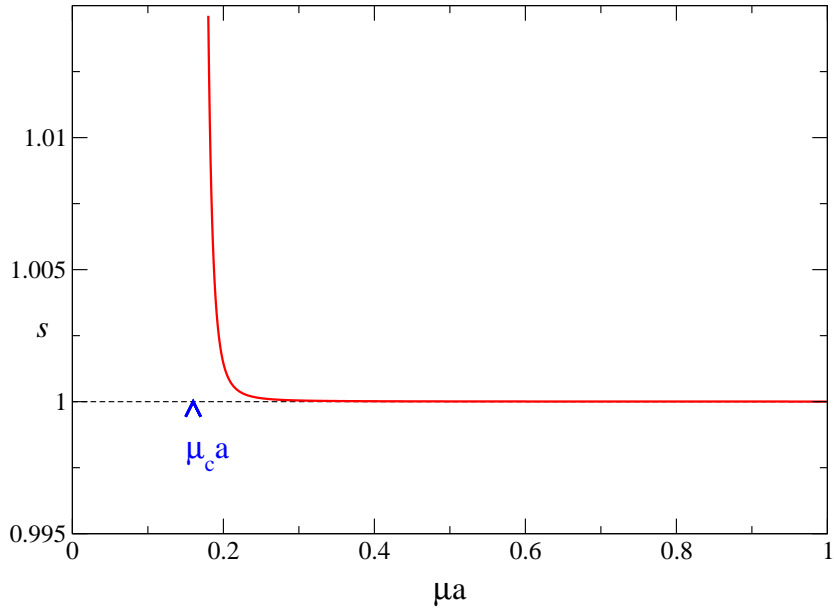}
\end{center}
\caption{The ratio $s=\beta_0/\beta_F$ as a function 
of $\mu a$ for the GN model with $N_f=4$} 
\label{fig:soundspeed}
\end{figure}

In Fig.~\ref{fig:soundspeed} we plot $s(\mu)$ for 
the model defined by ${\mathfrak g}=2$, $N_f=4$, $\mu_ca=0.16$ 
corresponding
to the simulation
results to be presented in the next section.
Note that chemical potential is expressed in
cutoff units. 
The plot only shows those solutions with $\mu a\gapprox0.18$
for which 
the physical sound speed 
$\beta_0=s\beta_F$ is definitely sub-luminal, using the relation
$\lim_{\mu\to\infty}\beta_F=1-{\mathfrak
g}/32N_f={63\over64}$ (note, though, that for finite $\mu$ 
this probably over-estimates $\beta_F$) \cite{HKTS}.
The plot shows $s$ to be a rapidly decreasing function of $\mu$, although 
the scale does not permit the observation that $\lim_{\mu\to\infty}
(s(\mu)-1)\simeq0.202\times10^{-5}>0$.

\begin{figure}[htb]
\bigskip\bigskip
\begin{center}
\epsfig{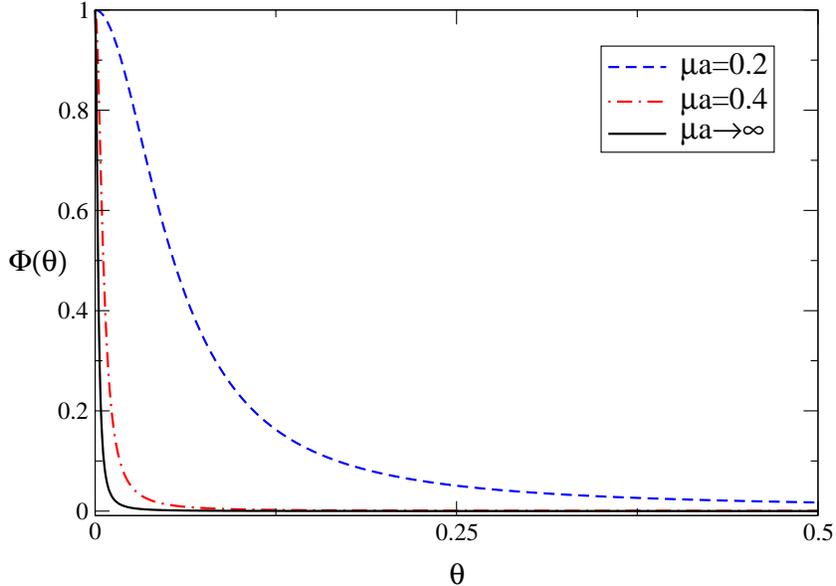}
\end{center}
\caption{Zero sound wavefunction $\Phi(\theta)$ for representative values
of $\mu a$ for the GN model with $N_f=4$.}
\label{fig:phi}
\end{figure}

To illustrate the form of the solution 
we plot $\Phi(\theta)$ for various $\mu$ in Fig.~\ref{fig:phi}.
The wavefunction has nodes at $\theta=\pm{\pi\over2}$, and
is highly peaked in the forward direction, once again 
due to the causality restriction that 
$s\sim1$ implying the denominator of
(\ref{eq:phi}) is very small as $\theta\to0$. This would appear to be a
general feature of weakly-interacting relativistic systems: indeed,
Fig.~\ref{fig:phi} shows the peak sharpen as $\mu$ increases 
and hence
$G(\mu)$ decreases, until a constant profile
is reached as 
$\mu\to\infty$. 

\section{Numerical Results}
\label{sec:numbers}

We have simulated lattice-regularised versions of the GN model
in 2+1$d$ with both Z$_2$ and SU(2)$\otimes$SU(2) global chiral symmetries. 
The underlying lattice action uses staggered fermions: further
details of the formulation and simulation algorithm are given 
in \cite{Simon,HM}.
Most of our
results are from the Z$_2$ version, for which
unless otherwise stated the simulation
parameters are identical to those of our previous study \cite{HKTS}, namely
$N_f=4$ and $a/g^2=0.75$, corresponding to a physical fermion mass at $\mu=0$
of $\Sigma_0a\simeq0.17$. We have studied both $32^2\times48$ lattices (the
volume used exclusively in \cite{HKTS}) and $48^3$; the increased momentum
resolution offered by the latter has proved to be important.
All results are taken in the chirally-restored phase,
ie. with $\mu>\mu_c\simeq0.16a^{-1}$. Measured 
values for the baryon density in lattice units $a=1$ are 
given in Table~\ref{tab:n}, and are to be compared to the saturation value 
$n_B(\mu\to\infty)=N_f/2$.
\begin{table}[h]
\centering
\setlength{\tabcolsep}{1.5pc}
\begin{tabular}{|ll|}
\hline
$\mu$ & $n_B$   \\
\hline
0.2     & 0.0226(2)   \\
0.4  &  0.1040(2)   \\
0.5  &  0.1730(2)   \\
0.6  &  0.2788(2)  \\
0.8  &  0.6254(2)  \\
\hline
\end{tabular}
\caption{Baryon density $n_B=\langle\bar\psi\gamma_0\psi\rangle$ for
the Z$_2$ model with $g^{-2}=0.75$.}
\smallskip
\label{tab:n}
\end{table}

By analysing the decay of the appropriate correlation function with Euclidean
time, 
we have calculated the dispersion relations $E(k)$
for the spin-${1\over2}$ 
quasiparticle, which carries a baryon charge, and 
$\omega(k)$ for various meson states
of the form $\bar\psi\Gamma\psi$ which are best thought of as 
excitations formed from a particle-hole
pair. We have simulated $L_s^2\times48$ systems with $L_s=32$ and 48, 
with $\mu$ ranging from 0.2 to 0.6,
and have
measured
$E(k),\omega(k)$ for $\vec k=(k,0)$ with $k=0,2\pi/L_s,\ldots,\pi/2$.
Note that the staggered fermion action is only invariant under translations
of two lattice spacings, which restricts the space of accessible momenta.
Further details can be found in \cite{HKTS}.

\begin{figure}[htb]
\bigskip\bigskip
\begin{center}
\epsfig{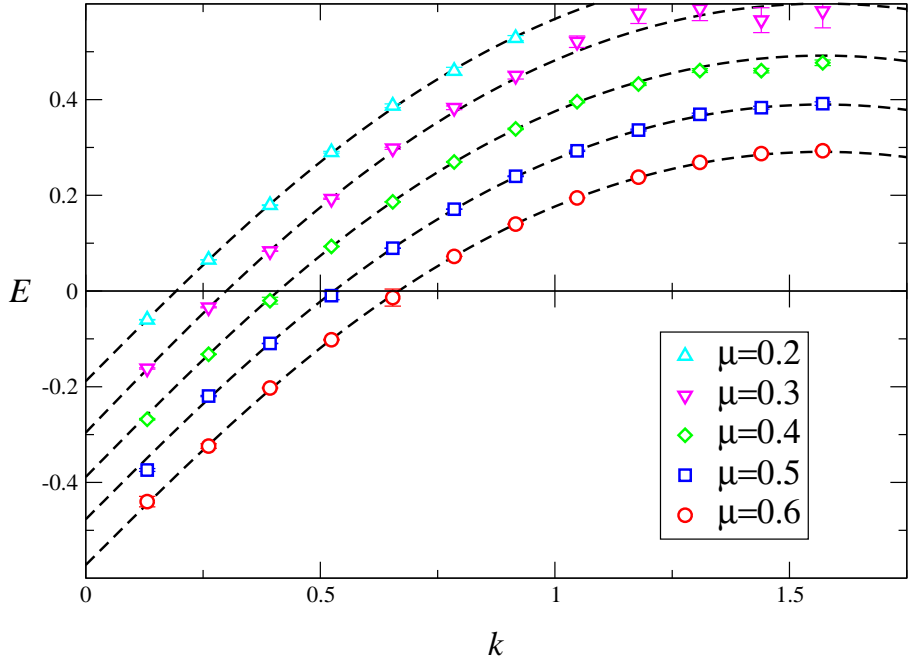}
\end{center}
\caption{Quasiparticle dispersion relation $E(k)$ from simulations
of the $Z_2$ GN model on $48^3$.}
\label{fig:ferm_disp}
\end{figure}
The quasiparticle propagator ${\cal G}(k,t)$
is measured using the formalism outlined in 
\cite{HLM}, although in this case an external diquark source term is not
introduced, so the ``anomalous'' propagator of the form
$\langle\psi(0)\psi^t(x)\rangle$ vanishes. In the chirally-restored phase
${\cal G}(k,t)$ vanishes for $t$ even; on the odd timeslices we fit to the
form
\begin{equation}
{\cal G}(k,t)=Ae^{-E_h(k)t}+Be^{-E_p(k)(L_t-t)}.
\end{equation}
On $48^3$
acceptable fits were found for the most part for $t\in[7,43]$, although the
fitting window had to be reduced somewhat for $\mu=0.2,0.3$. Empirically 
we find that for small $k$ the fit is dominated by a forward-propagating state
(ie. $A\gg B$) which is interpreted as a hole, and for large $k$ by a 
backward-propagating state ($A\ll B$) interpreted as a particle. The transition
between particle and hole behaviour is rather sudden, characteristic of a
well-defined Fermi surface. In Fig.~\ref{fig:ferm_disp} we show $E(k)$ with
hole energies $E_h$ plotted negative. Also shown are fits to the data with 
$k>0$ of the form
\begin{equation}
E(k)=-E_0+D\sinh^{-1}(\sin k),
\label{eq:disperse}
\end{equation}
ie. that of a Fermi liquid with discretisation effects taken into
account, with effective Fermi momentum $K_F$ and Fermi velocity $B_F$ given
by \cite{HLM}
\begin{equation}
K_F\equiv\sinh^{-1}(\sin k_F)={E_0\over D}\;\;\;;\;\;\;
B_F={{\partial\sinh(E+E_0)}\over{\partial\sin k}}\biggr\vert_{E=0}=
D{{\cosh E_0}\over{\cosh K_F}},
\label{eq:betaren}
\end{equation}
where $E(k_F)=0$.
The resulting $K_F$, $B_F$ and $K_F/\mu B_F$ are tabulated in 
Table~\ref{tab:fl}, and should be compared with the continuum 
values predicted in the
large-$N_f$ approach  for $\mu\gg\mu_c$ \cite{HKTS}:
\begin{equation}
\beta_F=1-{{\mathfrak g}\over{32N_f}}\simeq0.984\;\;\;;\;\;\;
{k_F\over{\mu\beta_F}}=1-{{\mathfrak g}\over{16N_f}}\simeq0.969.
\label{eq:fl1N}
\end{equation}
\begin{table}[htb]
\centering
\begin{tabular}{|llll|}
\hline
$\mu$ &  $K_F$ & $B_F$ & $K_F/\mu B_F$ \\
\hline
0.2 & 0.190(1) & 0.989(1) &  0.962(5) \\
0.3 & 0.291(1) & 1.018(1) &  0.952(4) \\
0.4 & 0.389(1) & 0.999(1) &  0.973(1) \\
0.5 & 0.485(1) & 0.980(1) &  0.990(2) \\
0.6 & 0.584(1) & 0.973(1) &  1.001(2) \\
\hline
\end{tabular}
\caption{Fermi liquid parameters resulting from fits of (\ref{eq:disperse})
to data from a $48^3$ lattice (quoted errors are purely statistical).}
\label{tab:fl}
\end{table}
The agreement is at best qualitative; the computed $O(1/N_f)$ corrections are 
so small that much better control over systematic effects would be 
required for a 
meaningful comparison with the data. For instance, 
the Fermi liquid parameters should be most accurately
pinned down for $\mu a\approx0.7$
implying roughly equal numbers of points with $k<k_F$ and $k>k_F$.
Note too that analytic prediction of $\beta_F$, $k_F$ to 
$O(1/N_f)$ for finite $\mu$ strictly
requires a knowledge of the gap equation to two loops, which is not yet
available \cite{HKTS}.
For the remaining analysis we 
will therefore assume the free field
values $K_F=\mu$, $B_F=1$. 

We have observed a similar situation in simulations of the SU(2)$\otimes$SU(2)
model on $32^2\times48$ with coupling $g^{-2}=0.85$ chosen so that the vacuum
fermion mass $\Sigma_0$ matches that of the Z$_2$ model at $g^{-2}=0.75$.
A much larger discrepancy in this model
between the measured Fermi liquid parameters and the prediction (\ref{eq:fl1N})
was reported in \cite{HLM}, probably due to the relatively small ratio
$(\mu-\mu_c)/\mu$ used in that study, implying that the Fermi liquid interaction
is strong and higher order effects are important.

\begin{figure}[htb]
\bigskip\bigskip
\begin{center}
\epsfig{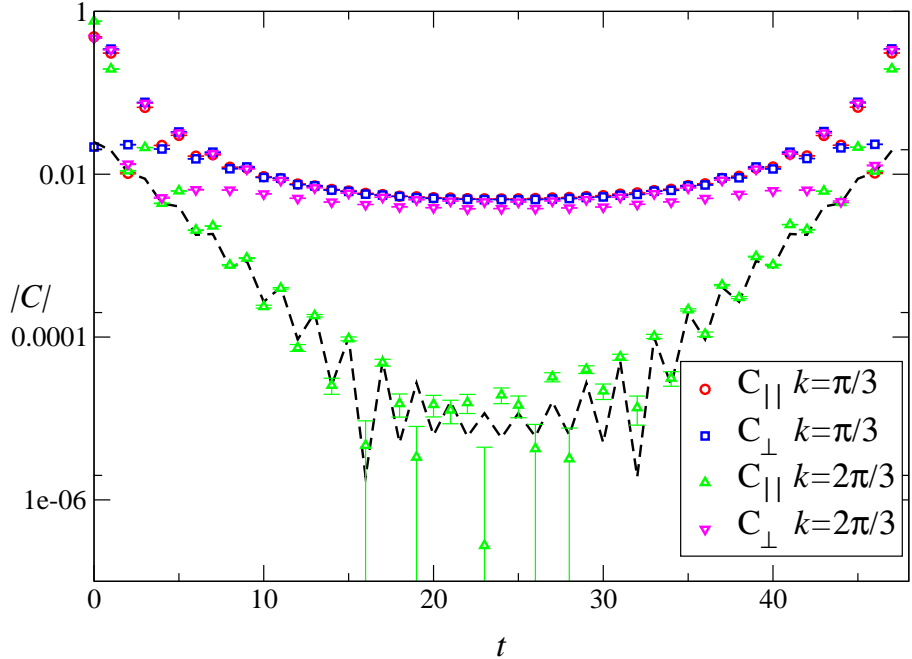}
\end{center}
\caption{Meson propagators $\vert C_\parallel\vert$ and $\vert C_\perp\vert$
for both $k={\pi\over3}$ and $\pi-{\pi\over3}$ for $\mu=0.5$. The dashed line
shows a fit of the form (\ref{eq:zsfit}).}
\label{fig:propagators}
\end{figure}
Next we consider the meson sector, ie. correlators of the form
\begin{equation}
C_\Gamma(\vec k,t)
\equiv\sum_{\vec x}
\langle \bar\psi\Gamma\psi(\vec0,0)\;\bar\psi\Gamma\psi(\vec x,t)\rangle
e^{i\vec k.\vec x}, 
\end{equation}
which carry
zero baryon number. In Ref.~\cite{HKTS} we showed that the generic 
large-distance behaviour
in any given channel is  dominated by zero-energy particle-hole pairs, and as a
result the decay is algebraic, ie. $C_\Gamma(t)\propto t^{-\lambda(\vec k)}$, 
with the
exponent $\lambda$ determined by the geometry of the overlap between two Fermi
disks with relative displacement $\vec k$ between their centres.
A particularly favourable configuration occurs for $\vert\vec
k\vert\approx2\mu$, in which case the Fermi surfaces just kiss, 
and $\lambda={3\over2}$. In this paper we focus on point-split
meson operators corresponding to spatial components of the conserved vector
current, ie. of the form
\begin{equation}
j_i(x)={\eta_i(x)\over2}\left[\bar\chi(x)\chi(x+\hat\imath)+\bar\chi(x+\hat
\imath)\chi(x)\right],
\end{equation}
where $\chi,\bar\chi$ are staggered fermion fields, $\eta_1(x)=(-1)^{t}$,
and $\eta_2(x)=(-1)^{t+x_1}$. Just as before we set $\vec k=(k,0)$,
and define $C_\parallel$ in terms of the correlator $\langle
j_1(0)j_1(x)\rangle$ and $C_\perp\sim\langle j_2(0)j_2(x)\rangle$. In
Fig.~\ref{fig:propagators} we show $\vert C_{\parallel,\perp}\vert(k,t)$ 
data taken on a $48^3$ lattice 
at $\mu=0.5$, for both
$k={\pi\over3}\approx2\mu$ and $k=\pi-{\pi\over3}$ (the modulus is taken
because we use a log-linear scale, and $C$ fluctuates in sign). Whereas 
$C_\perp$ and $C_\parallel(k={\pi\over3})$ all show the expected algebraic
decay, the decay in the $C_\parallel(k=\pi-{\pi\over3})$ channel is much 
faster, and resembles the exponential decay expected of an isolated pole.
We have fitted it with the form
\begin{equation}
C_\parallel(\pi-k,t)=A\exp(-\Omega(k)t)+(-1)^tB\exp(-\omega(k)t),
\label{eq:zsfit}
\end{equation}
in most cases employing datapoints with $t\in[10,38]$.

\begin{figure}[htb]
\bigskip\bigskip
\begin{center}
\epsfig{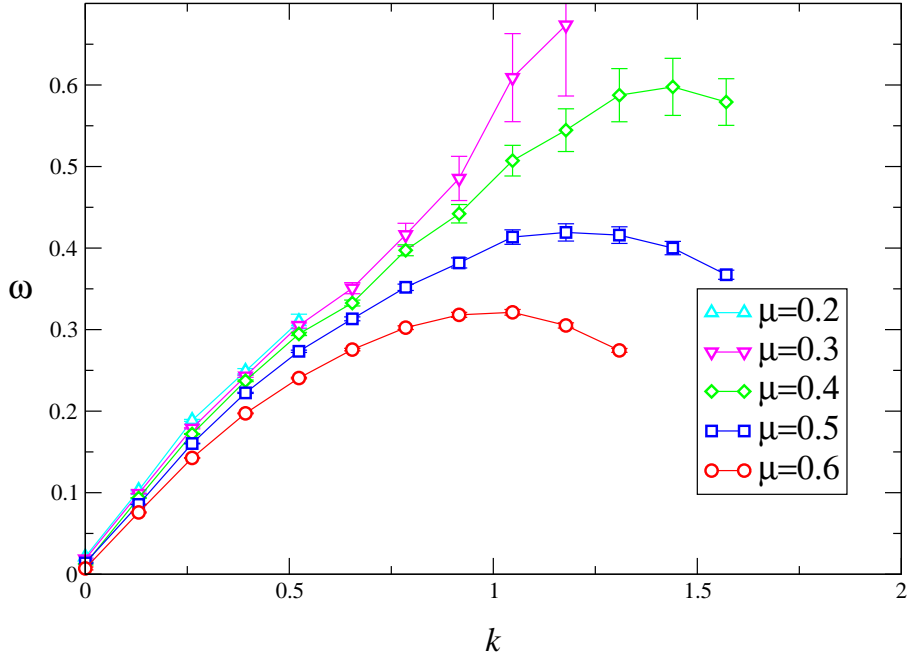}
\end{center}
\caption{Dispersion relation $\omega(k)$ for various $\mu$ for the 
channel defined by the correlator
$C_\parallel(\pi-k)$. Data are from $48^3$.}
\label{fig:rho_disp}
\end{figure}
For small $k$, the coefficient $B\gg A$, suggesting that the correlator is
dominated by a pole in the alternating channel. In Fig.~\ref{fig:rho_disp}
we plot the resulting dispersion relation $\omega(k)$ for $\mu=0.2,\ldots,0.6$.
The behaviour $\omega\propto k$ as $k\to0$ suggests the presence of a massless 
pole similar to that of a phonon. Support for this interpretation
comes from recasting the meson bilinear in terms of continuum-like fields
$q^{\alpha a}$, $\bar q^{\alpha a}$ having spinor index $\alpha=1,\ldots4$ and
`color' index $a=1,2$ \cite{BB}. We obtain
\begin{equation}
(-1)^{x_1}(-1)^t\bar\chi(x)\eta_1\chi(x+\hat1)\sim i\bar
q(\gamma_0\otimes\tau_2^*)q,
\end{equation}
demonstrating that the excitation can be viewed 
as an oscillation of local baryon density.

Since we are attempting to simulate low temperatures $T\ll\mu$, 
our expectation
is that the phonons are characteristic of zero sound and should be regarded as
collective excitations of the degenerate ground state, rather than 
of first sound which relies on particle collisions to generate the
elasticity of the medium, and which is therefore heavily damped for
small $T$ \cite{Landau2}. An obvious objection to this proposal is that the 
phase velocity $\beta_0=\omega/k$ 
derived from Fig.~\ref{fig:rho_disp} has
$\beta_0\sim0.55$ - 0.65, whereas the theoretical considerations of
Sec.~\ref{sec:theory} imply that undamped oscillations in relativistic
matter should have $\beta_0\approx1$. However, recall
that although we are examining the $k\to0$ limit, we are trying to model a
Fermi
surface phenomenon on the lattice; 
the appropriate scale with which to compare $\beta_0$
is the ``bare'' Fermi velocity (ie. without the discretisation
correction modelled by (\ref{eq:disperse},\ref{eq:betaren}), which for free
fields is given by
\begin{equation}
\beta_F^{\rm bare}={{\partial E}\over{\partial k}}\biggr\vert_{E=0}=
\sqrt{{1-\sinh^2\mu}\over{1+\sinh^2\mu}}.
\end{equation}
\begin{figure}[htb]
\bigskip\bigskip
\begin{center}
\epsfig{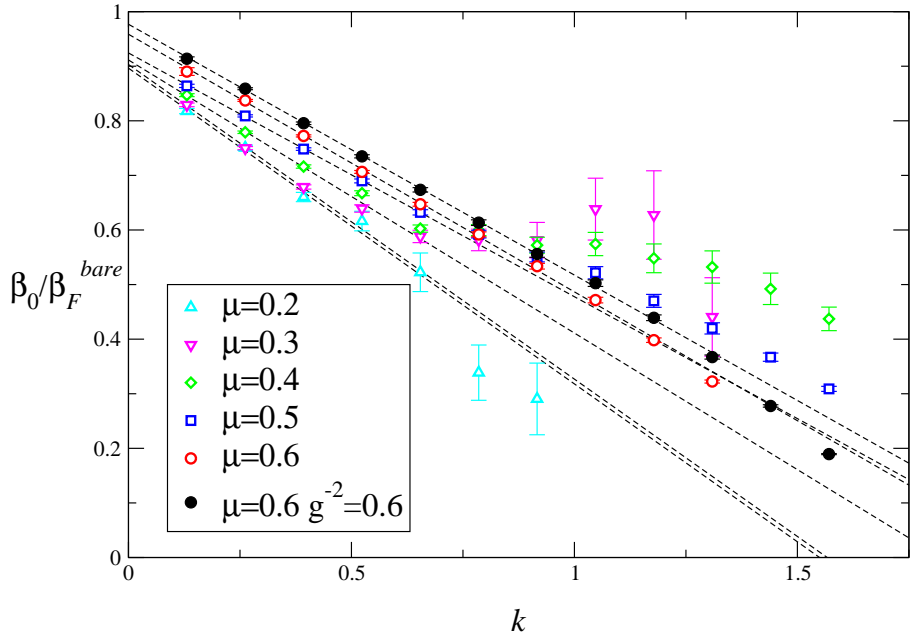}
\end{center}
\caption{Sound velocity $\beta_0$ rescaled by the bare Fermi velocity
$\beta_F^{\rm bare}$ for various $\mu$.}
\label{fig:zerosound}
\end{figure}

In Fig.~\ref{fig:zerosound} we plot $\beta_0/\beta_F^{\rm bare}$ for $k>0$
using the $48^3$
data of Fig.~\ref{fig:rho_disp}. There are two striking features. First, the
ratio appears to extrapolate to a value rather closer to, though
still smaller than, 1 as $k\to0$, getting closer as $\mu$ increases until
for $\mu=0.6$ the intercept $>0.95$. Secondly, for $k\leq{{5\pi}\over24}$
the curves are remarkably close to straight lines, 
implying a dispersion of the form $\omega=sk-Fk^2$. Fig.~\ref{fig:zerosound}
also shows fits of this form, and the resulting intercept and slope are
given for all the systems studied in Table~\ref{tab:speeds}.
\begin{table}[htb]
\centering
\begin{tabular}{|l|ll|ll|}
\hline
 & $48^3$ &  & $32^2\times48$ & \\
$\mu$ &  $s$ & $F$ & $s$ & $F$ \\
\hline
\hline
Z$_2$ $(g^{-2}=0.75)$    &&  &  & \\
0.2 & 0.896(8) & 0.578(4) &  - & -  \\
0.3 & 0.903(5) & 0.576(20) & - & - \\
0.4 & 0.911(4) & 0.500(16) & 0.894(5) & 0.552(15) \\
0.5 & 0.924(3) & 0.447(9) & 0.961(3) & 0.575(8) \\
0.6 & 0.959(2) & 0.472(4) & 0.950(2) & 0.425(2) \\
0.7 &  -       & -        & 1.033(3) & 0.556(8) \\
\hline
\hline
Z$_2$ $(g^{-2}=0.6)$ &&&&\\
0.6 & 0.977(2) & 0.460(5) & - & - \\
\hline
\hline
SU(2)$\otimes$SU(2) $(g^{-2}=0.85)$ &&&&\\
0.5 $\;\;\;\;\;I_3=1$ & - & - & 0.938(2) & 0.467(7) \\
0.5 $\;\;\;\;\;I_3=0$ & - & - & 0.917(3) & 0.453(9) \\
\hline
\end{tabular}
\caption{Straight line fits to $\omega/k\beta_F^{\rm bare}$ vs. $k$ 
(quoted errors are purely statistical).}
\label{tab:speeds}
\end{table}
The most successful fits were to data on $48^3$ with $\mu=0.6$, where 
straight line fits to data with $k\leq{{3\pi}\over8}$ proved possible.
For $\mu\leq0.4$ the fits only included data with $k\leq{\pi\over8}$ ($48^3$)
or $k\leq{{3\pi}\over16}$ ($32^2\times48$). The general trend revealed in 
Table~\ref{tab:speeds} is that the speed ratio $s$ increases towards unity
as $\mu$ increases, 
and that the slope parameter $F$ decreases slightly.
Since the momentum resolution on $32^2\times48$ is markedly worse, we can only 
comment on the absence of evidence for any significant finite volume effects, 
and, in the  model with global SU(2)$\otimes$SU(2) symmetry, for any significant
difference between isosinglet (`$\omega$') and isotriplet (`$\rho$') channels.

The most interesting comparison is between Z$_2$ data on $48^3$ with
$g^{-2}=0.6$ and 0.75, the stronger coupling corresponding to 
roughly
a factor of two increase in lattice spacing; if we restore explicit factors
of $a$ we see that the slope parameter $Fa^{-1}$ remains roughly constant
under this change, implying that the slope is a lattice artifact which should
vanish in the continuum limit. At first sight, it is a little strange to observe
an $O(a)$ scaling violation in a simulation with staggered fermions. However,
a careful analysis \cite{HKK} reveals that there are indeed $O(a)$
discretisation artifacts in the lattice formulation of the GN interaction term
$(\bar\psi\psi)^2$. This highlights that the effect we are examining is 
a property of an interacting system, since non-interacting staggered fermions
show discretisation effects only at $O(a^2)$. Since $s$ also increases as the 
coupling is strengthened we deduce a systematic effect as $\mu a$ and hence 
$\mu/T$ increase.

\section{Discussion}
\label{sec:discuss}

We have, in Sec.~\ref{sec:theory}, used the Fermi liquid interaction to
leading non-trivial order in $1/N_f$ to find solutions to the Boltzmann
equation corresponding to zero sound. The excitation is spin and 
isospin symmetric, and has speed $s=\beta_0/\beta_F>1$ for almost all 
$\mu>\mu_c$. In the simulations described in 
Sec.~\ref{sec:numbers} we have found a massless pole
in a particular mesonic channel, interpolated by a staggered fermion bilinear
point-split along the direction of propagation, with continuum
quantum numbers corresponding to a local
fluctuation in baryon density, with speed $s\lapprox1$.
To what extent can we be
sure these results describe the same physical phenomenon?

Our solution based on the ansatz (\ref{eq:phi}), is the unique one
(symmetric in internal quantum numbers) which is a meromorphic function of
$z=e^{i\theta}$. 
As remarked above, the
poles fall on the real axis, yielding a real $s>1$.  Solutions with
$s<1$ must therefore be of a different form to (\ref{eq:phi}), and probably
involve a branch cut ending inside the $\vert z\vert=1$ contour. We can see
this physically by considering the emission of a phonon with momentum $\vec q$
and energy $\beta_0\vert\vec q\vert$ from a quasiparticle with momentum
$\vec k$ and energy $\mu+\beta_F(\vert\vec k\vert-k_F)$. 
This is allowed kinematically for $s<1$, the angle of 
emission $\phi$ satisfying
\begin{equation}
\cos\phi=s+{{\vert\vec q\vert}\over{2\vert\vec k\vert}}(1-s^2)>s.
\end{equation}
All radiation is emitted within a cone of half-angle $\cos^{-1}s$ centred on
the quasiparticle trajectory. This well-known phenomenon is variously known
as Landau damping, \v Cerenkov radiation, or most
appropriately in the current context, as a sonic boom. 

On a spacetime lattice, this process is constrained because there is a
natural lower bound for the angle of emission, $\phi_{\rm
min}\sim2\pi/L_s\vert\vec q\vert$. Landau damping is thus kinematically 
forbidden for
\begin{equation}
s>\cos\phi_{\rm min}\simeq1-{{2\pi}\over{L_s^2\vert\vec q\vert^2}}.
\end{equation}
With a conservatively high $\vert\vec q\vert\sim\mu/2$, then on 
a $48^3$ lattice we have no damping for $s>0.73$ at $\mu=0.2$, rising to
$s>0.97$ at $\mu=0.6$. It is plausible therefore that 
damping is suppressed and that the phonon found in
Sec.~\ref{sec:numbers} is described by an isolated
pole even if $s<1$; indeed, Fig.~\ref{fig:zerosound}
suggests that $s$ increases towards one as $\mu$ increases.
This contradicts the analytic solution shown in Fig.~\ref{fig:soundspeed}, 
which has $\lim_{\mu\to\infty}s\to1_+$; note, however, the disparity in vertical
scale between the two figures. 
Simulations on volumes 
considerably larger than those used here will be needed to disentangle the
various possible systematic effects due to finite $L_s$, finite $\mu/T$, and
non-zero $a$ in order to determine the sign of $s-1$ for finite $N_f$.
In principle, since our extracted value of $s\lapprox1$ 
has relied on a rescaling
of $\beta_F$ to take account of discretisation artifacts, we 
also need to go to considerably 
finer lattices in order to demonstrate a clear distinction
between our signal and first sound with expected
propagation speed $\beta_1\simeq0.7$.
If, however, the theoretical arguments about zero sound dominating as $T\to0$
can be taken seriously, then for the 
first time we have succeeded 
in identifying a collective oscillation in a lattice
simulation.

\section*{Acknowledgements}
SJH was supported by a PPARC Senior Research Fellowship, and
thanks the Institute for Nuclear Theory at the University of Washington for its
hospitality and the Department of Energy for partial support during the
completion of this work.

\end{document}